\documentclass[a4paper,fleqn,usenatbib]{mnras}
\DeclareMathAlphabet{\mathpzc}{OT1}{pzc}{m}{it}
\twocolumn
\usepackage{mathptmx}

\usepackage[T1]{fontenc}
\usepackage{ae,aecompl}

\usepackage{graphicx}	
\usepackage{amsmath}	
\usepackage{amssymb}	
\usepackage{bm}
\usepackage{xcolor}

\title[Large-scale structures in the $\Lambda$CDM Universe]{Large-scale structures in the $\Lambda$CDM Universe:\\ network analysis and machine learning}

\author[M. Tsizh et al.]{
Maksym Tsizh $^1$,\thanks{E-mail: maksym.tsizh@lnu.edu.ua}
Bohdan Novosyadlyj $^{1,2}$, Yurij Holovatch$^{3,4,5}$,
Noam I Libeskind $^{6,7}$
\\
$^1$Ivan Franko National University of Lviv, Kyryla i Methodia str. 8, Lviv, 
79005, Ukraine \\
$^2$ College of Physics and International Center of Future Science of Jilin University,
Qianjin str. 2699, Changchun, 130012, P.R.China \\
$^3$Institute for Condensed Matter Physics, National Acad. Sci. of Ukraine, 79011 Lviv, Ukraine \\
$^4$ $\mathbb{L}^4$ Collaboration \& Doctoral College for the Statistical Physics of Complex Systems, Leipzig-Lorraine-Lviv-Coventry, Europe \\
$^5$Centre for Fluid and Complex Systems, Coventry University, Coventry, CV1 5FB, United Kingdom\\
$^{6}$Leibniz-Institut f\"ur Astrophysik Potsdam (AIP),  An der Sternwarte 16, 14482 Potsdam, Germany\\
$^{7}$University of Lyon, UCB Lyon 1, CNRS/IN2P3, IPN Lyon, France\\
}
\date{Accepted XXX. Received YYY; in original form ZZZ}

\pubyear{2019}
\begin{document}
\label{firstpage}
\pagerange{\pageref{firstpage}--\pageref{lastpage}}
\maketitle

\begin{abstract}
We perform an analysis of the Cosmic Web as a complex network, which is built on a $\Lambda$CDM cosmological simulation. For each of nodes, which are in this case dark matter halos formed in the simulation, we compute 10 network metrics, which characterize the role and position of a node in the network. The relation of these metrics to topological affiliation of the halo, i.e. to the type of large scale structure, which it belongs to, is then investigated. In particular, the correlation coefficients between network metrics and topology classes are computed. We have applied different machine learning methods to test the predictive power of obtained network metrics and to check if one could use network analysis as a tool for establishing topology of the large scale structure of the Universe. Results of such predictions, combined in the confusion matrix, show that it is not possible to give a good prediction of the topology of Cosmic Web (score is $\approx$ 70 $\%$ in average) based only on coordinates and velocities of nodes (halos), yet network metrics can give a hint about the topological landscape of matter distribution.
\end{abstract}

\begin{keywords}
cosmology: large scale structure--gravitationally bound systems--complex networks-machine learning
\end{keywords}

\section{Introduction}

Applying complex network methods is one of the latest trends in studying of the large scale structure of the Universe. This approach is the next step after excursion set theory of halos \cite{Bond91, ShethTormen2002} and voids \citep{ShethWeygaert2004} formation and evolution in cosmology. Unlike the excursion set theory, in the complex network approach, the hierarchy of nodes is not a crucial feature. However, the importance for connectivity of network, place in the network, neighbour richness and centralities of the nodes are.

The pioneering work of this approach in cosmology \citep{HongDei15} addressed the problem of relation between network centralities of nodes in the Cosmic Web and the type of topological population of the corresponding galaxy. The following papers studied different usages of complex networks analysis of the Cosmic Web: discriminating of different topologies in population with similar two-point correlation functions \citep{Hong16}, discovering various ways of network construction \citep{Coutinho16}, finding similarity and peculiarity of physical galaxy properties (color, brightness, mass index) of different topology (defined by network characteristic) environment population \citep{Apunevych17}, relating correlation function and relative size of the largest connected component of network \cite{Zhang18}, studying connectivity of Gaussian Random Field-like galaxy field  \citep{Codis18} as a probe of evolution of structures and the nature of dark energy. In recent works, \citep{HongJeong19} and \citep{HongDey19}, authors study how the transitivity of the Cosmic Web and its other network characteristic can distinguish models of dark energy in cosmological simulations and different scenarios  of Ly-alpha emitters.
Aside the complex network approach there are other graph-based methods of studying the Cosmic Web. One of them is the minimal spanning trees (MST) approach. It was successively developed in a number of papers \citep{Barrow85, Graham95, Colberg07}. One of the latest works in this subject exploits the MST to identify and classify large-scale structures (filaments and voids) within the Galaxy And Mass Assembly (GAMA) survey \citep{Alpaslan13}.

We would like also to mention investigations of the topological features of the Cosmic Web through Betti numbers \citep{Pratyush16} and beta-skeleton analysis \citep{Fang18}, both also used to reveal the underlying structures in galaxy distribution.

Another interesting and rapidly developing direction in data analysis of the large scale structure and extra-galactic astronomy is the usage of machine learning (ML) for object classification or their parameters and features estimation. Wide review of this field of research is beyond the scope of this paper, however, one could easily find dozen of papers in the recent years. For example: morphological classification of galaxies by their images \citep{Huertas07} or their observable features \citep{Dobrycheva17}; predicting galaxy features like HI content optical data \citep{Rafieferantsoa18, Zamudio19} or multi-wavelength counterparts of sub-millimeter galaxies \citep{An18} or galaxy cluster mass \citep{Armitage18} based on surveys. ML methods can paint galaxies themselves, knowing only information about host dark matter halo \citep{Agarwal17, Zhang19}. Finally, it is possible to predict directly the evolution of cosmological structure formation (in terms of Press-Schechter theory) \citep{Lucie18,He18} or simulate the Cosmic Web \citep{Rodriguez18} or weak lensing map \citep{Mustafa17} via ML . 

In this work, for the first time, we combine both powerful methods (complex network analysis and ML) with a purpose to test whether such combination can become a new tool of probing the topology of large scale structure. We choose to use the results of GADGET2 cosmological simulation \citep{Libeskind17}, benefiting from the fact that 12 well established structure finders have already been applied to it. 

This work has the following structure. In the next section we describe in details the construction of the network and topology classification of the Cosmic Web. In the third section we define all the network metrics which we are going to use and show their correlations with topology structure types and between themselves. Also, we provide distributions of the network metrics for different structure sub-populations. In the fourth section we will use ML methods to predict the type of topology structure to which each halo belongs, having the very minimum information about it (coordinate, mass, spin and velocity) by utilizing computed network metrics as predictors for ML. We discuss the obtained results in the final section.

\section{Building a network}
\subsection{Network on $\Lambda$CDM Universe}
In this paper we rely on data of cosmological simulation GADGET2 performed and provided by N. Libeskind and co-authors in work \citep{Libeskind17} to build the network of the Cosmic Web. This simulation is one-type particle cosmological $N$-body simulation of dark matter distribution and assumes following $\Lambda$CDM cosmological parameters: $h=0.68$, $\Omega_M = 0.31$, $\Omega_\Lambda = 0.69$, $n_s = 0.96$, and $\sigma_8 = 0.82$. The size of the box is 200 $h^{-1}$ Mpc with $512^3$ dark matter particles of mass $5.12\cdot 10^{9}$ $M_\odot/h$ in it. Haloes in the simulation were identified by a friends-of-friends (FOF) algorithm, with a linking length of $0.2\times$ (mean particle separation) $h^{-1}$ Mpc and a minimum of 20 particles per halo.  The result was a catalogue of 281465 halos with mass range $10^{11}$ -- $10^{15}$ $h^{-1}$ $M_\odot$.  

We start by representing the set of halos as a complex network. Each halo in the catalogue is a node of network. The nodes with distances between them smaller than a certain value, called linking length $l$, are connected by edges. Note, that this linking length is not be confused with the one that unites particles in halos. There can be only 1 or 0 edges between two nodes (network has no multiplicity). Here the simplest possible option is considered: the edges are unweighted and undirected. As a result simple undirected unweighted graph is built, which we analyse. Such graph can be naturally described by its symmetric adjacency matrix  $\mathbf{A}$, in which its element $a_{ij}$ is 1 when $i$-th and $j$-th nodes are connected and 0 otherwise, $a_{ii}=0$.
\subsection{Linking length choice}
\begin{figure*}
  \begin{center}
  \includegraphics[width=0.48\textwidth]{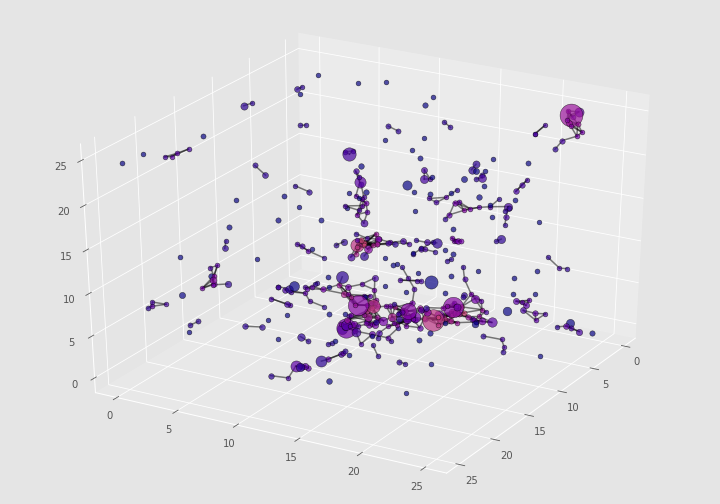}
  \includegraphics[width=0.48\textwidth]{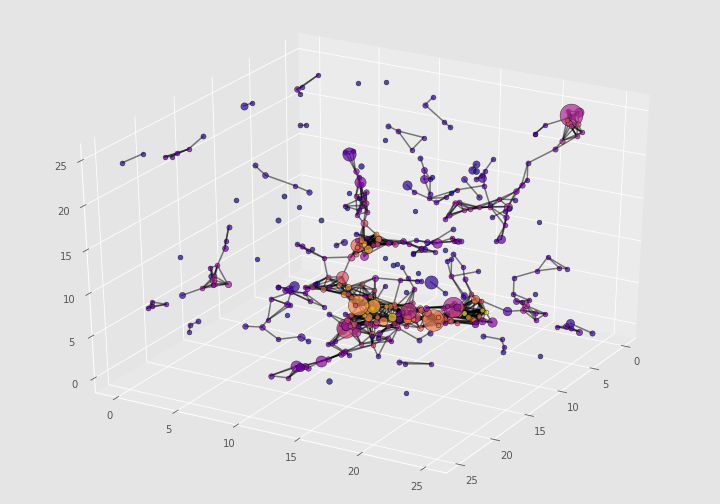}
  \end{center}
  \caption{A 25x25x25 Mpc/h part of the simulation cube with halos and links between them for $l=1.6$ Mpc/h(on the left) and $l=2.4$ Mpc/h (on the right). Brighter color of halos denotes higher number of neighbours. The sizes are proportional to halos' mass}
\label{figll}
\end{figure*}
\begin{figure}
  \begin{center}
  \includegraphics[width=0.48\textwidth]{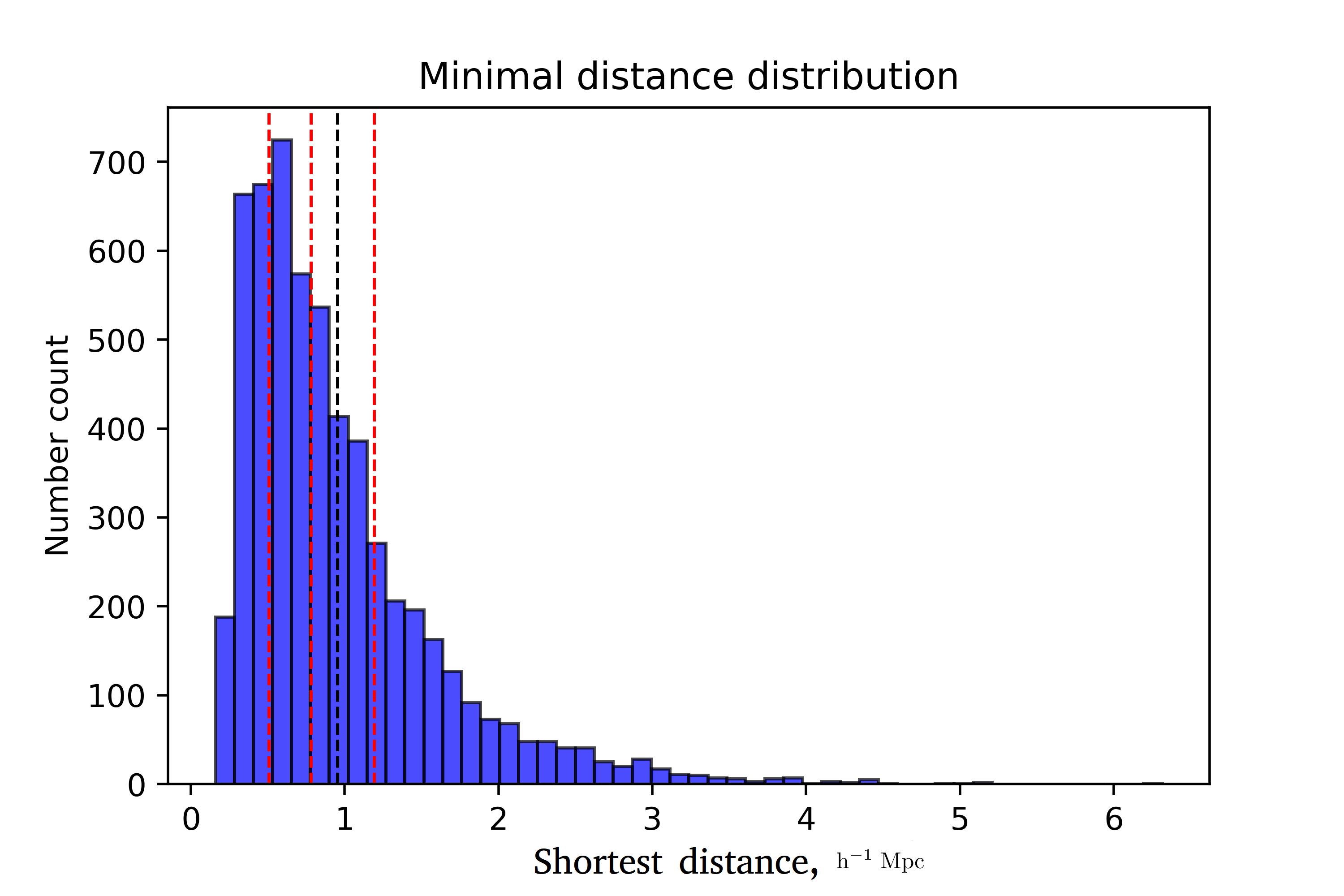}
  \end{center}
  \caption{Distribution of distances to the closest neighbour for sample of population. Red lines denotes first, second and third quartiles, black line denote mean value.}
  \label{fig1a}
\end{figure}

As it has been noted before \citep{HongDei15,Apunevych17}, network linking length affects all the numbers, relations and laws one may discover when exploring the network. Indeed, having linking length too small will result in the disconnected network and reveal no structures in the web, while having it too large will not show the peculiarities and  important nodes of the network. Nevertheless, there is a range of linking length values, for which properties of constructed networks are similar. We have experimented with the range of $l$ between 1.6 and 2.4 $h^{-1}$ Mpc in 0.1Mpch $h^{-1}$ intervals (see the difference between networks with different linking length in Fig.\ref{figll}). Though the average values of metrics vary with linking length, the computed correlations we discuss here have shown qualitatively similar behaviour within all the range of values of linking length. For the detailed analysis of the network here, we choose the linking length to be approximately twice larger than average distance to the closest neighbour, $l=2$ $h^{-1}$ Mpc $= 2.94$ Mpc (see Figure \ref{fig1a}). It must be said, that there is no direct hint of how large the linking length should be, there is no any special scale in analysing such spatially diverse topological structures. Yet we believe, that the range we chose is inside scales that are suitable for analysis, as the characteristics of resulting network are similar to those analysed in other works.

Indeed, in terms of $l_{\bar{\mu}}$ introduced in \cite{HongDei15} our linking length $1.6 Mpc/h < l < 2.4 Mpc/h$ is less than $l_2$, while those authors use $l_5$ and $l_6$ in their work. In this notation $l_\mu$ is such linking length, that in random network with same number of nodes mean value of nodes inside radius $l_\mu$ is $\mu$ (formula (25) in \cite{HongDei15}). However, in our opinion, the crucial parameters are average values of metrics, degree $k$ and clustering coefficient $Cl$ (definitions are given in the next section) in particular. These values at $l=2$ $h^{-1}$ Mpc are close to ones computed in \cite{Apunevych17} in their network analysis, $\langle k \rangle = 6.77$, $\langle Cl \rangle = 0.603$ (see the Table 1 for average values we obtained in our networks). Therefore, we conclude that chosen value is appropriate to study the Cosmic Web we consider.

\begin{table}
\centering
\begin{tabular}{||c c c c c||} 
 \hline
 Cosmic Web/ &\multicolumn{3}{| c |}{$l$ in [$h^{-1} $Mpc]}\\
 Metrics & 1.6 & 2.0 & 2.4 \\ [0.5ex] 
 \hline
 $\langle k \rangle$ & 3.5 & 5.6 & 8.1 \\ 
 \hline
 $\langle A_n\rangle$ & 3.9 & 6.1 & 8.8 \\
 \hline
 $\langle C_{\mathrm{K}} \rangle$ & 1.87 & 1.86$\cdot 10^{-3}$ & 1.72$\cdot 10^{-3}$ \\
 \hline
 $\langle C_b \rangle$ & 8.7 & 7.8 & 0.0002 \\
 \hline
 $\langle C_c\rangle$ & 0.04$\cdot 10^{-3}$ & 0.88$\cdot 10^{-3}$ & 5.25$\cdot 10^{-3}$ \\
 \hline
 $\langle C_h\rangle$ & 18.9 & 351.8 & 1827 \\
 \hline
 $\langle T \rangle$ & 6.9 & 17.7 & 37.7 \\
 \hline
 $\langle Cl \rangle$ & 0.41 & 0.51 & 0.56 \\
 \hline
 $\langle Sq \rangle$ & 0.22 & 0.25 & 0.24 \\
 \hline
 $\langle C_x \rangle$ & 6.6$\cdot 10^{-5}$ & 6.8$\cdot 10^{-5}$ & 6.3$\cdot 10^{-5}$ \\
\end{tabular}
\caption{Average values of network metrics (node degree $\langle k \rangle$, average neighbour degree $\langle A_n \rangle$, Katz centrality $\langle C_{K}\rangle$,  betweenness centrality $\langle C_b \rangle$, closeness centrality $\langle C_c\rangle$, harmonic centrality $\langle  C_c\rangle$, triangles  $\langle T \rangle$, clustering coefficient $\langle Cl \rangle$, squares  $\langle Sq \rangle$ and eigen centrality  $\langle C_x \rangle$) for networks with different linking length $l$. Find the definitions of these metrics in the following section.}
\label{table1}
\end{table}
\subsection{Topology classification of the large scale structures}
The 12 methods of topology classification of the large scale structure compared in \cite{Libeskind17}  were used to define the affiliation of each halo to up to 4 possible types of structures: voids, filaments, sheets and knots (superclusters). These classifications and their link to network characteristic of each halo are of the main interest of this work.

\begin{figure}
  \begin{center}  
  \includegraphics[width=0.48\textwidth]{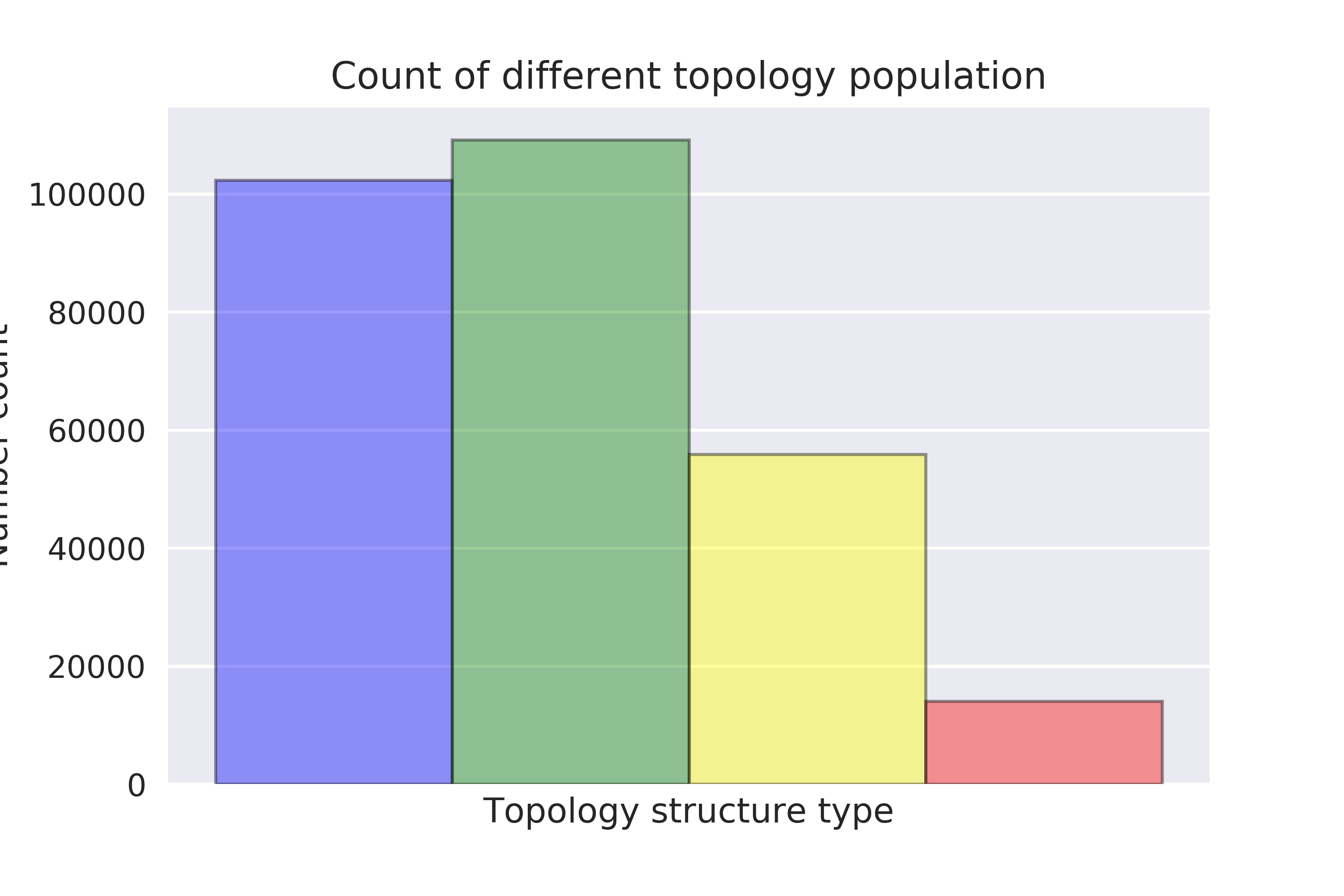}
  \end{center}
  \caption{Number count of population of different topology classes  - voids (blue), filaments (green), sheets (yellow) and knots (red). Filaments are most populated.}
  \label{fig1b}
\end{figure}

\begin{figure*}
  \begin{center}
  \includegraphics[width=0.48\textwidth]{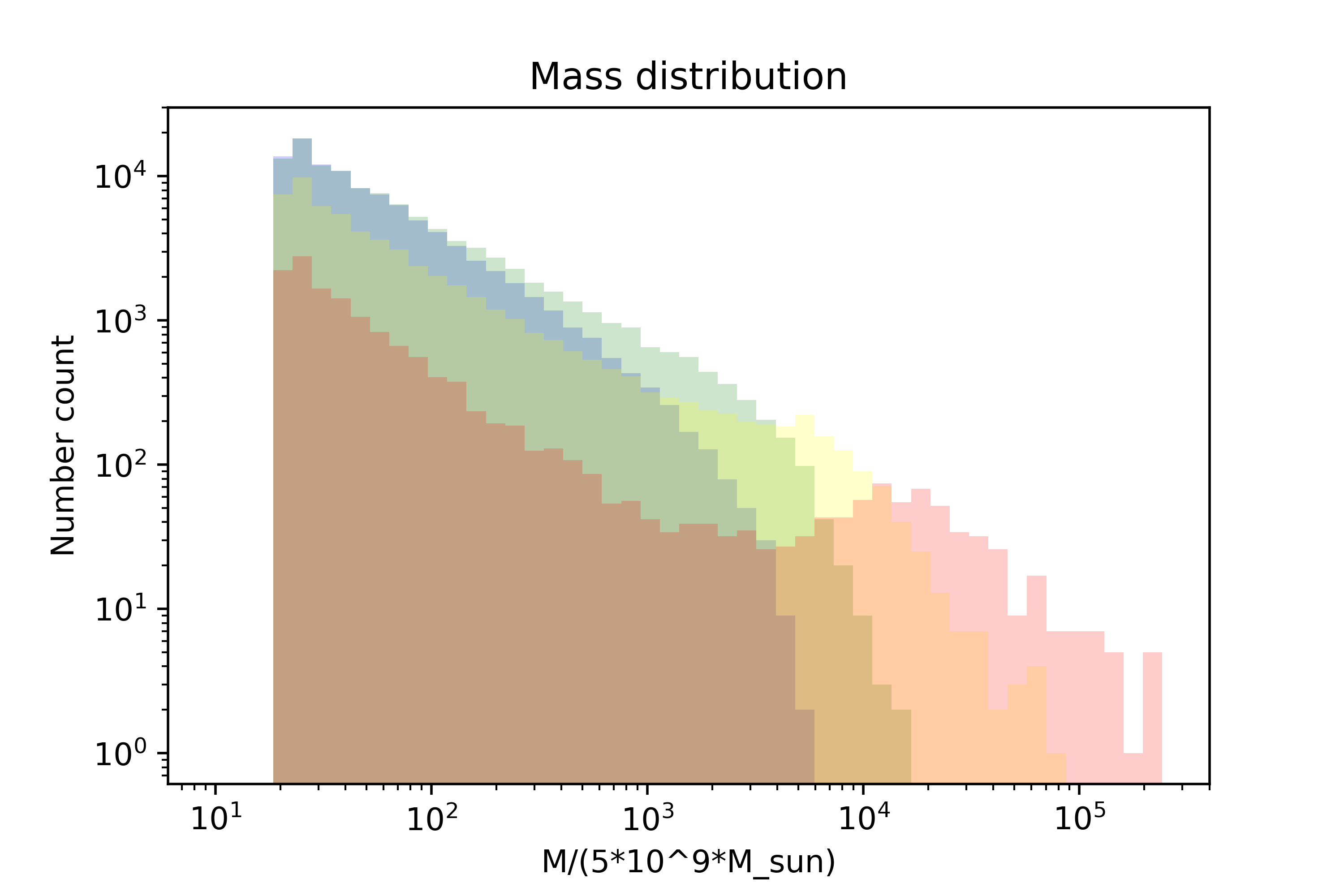}
  \includegraphics[width=0.48\textwidth]{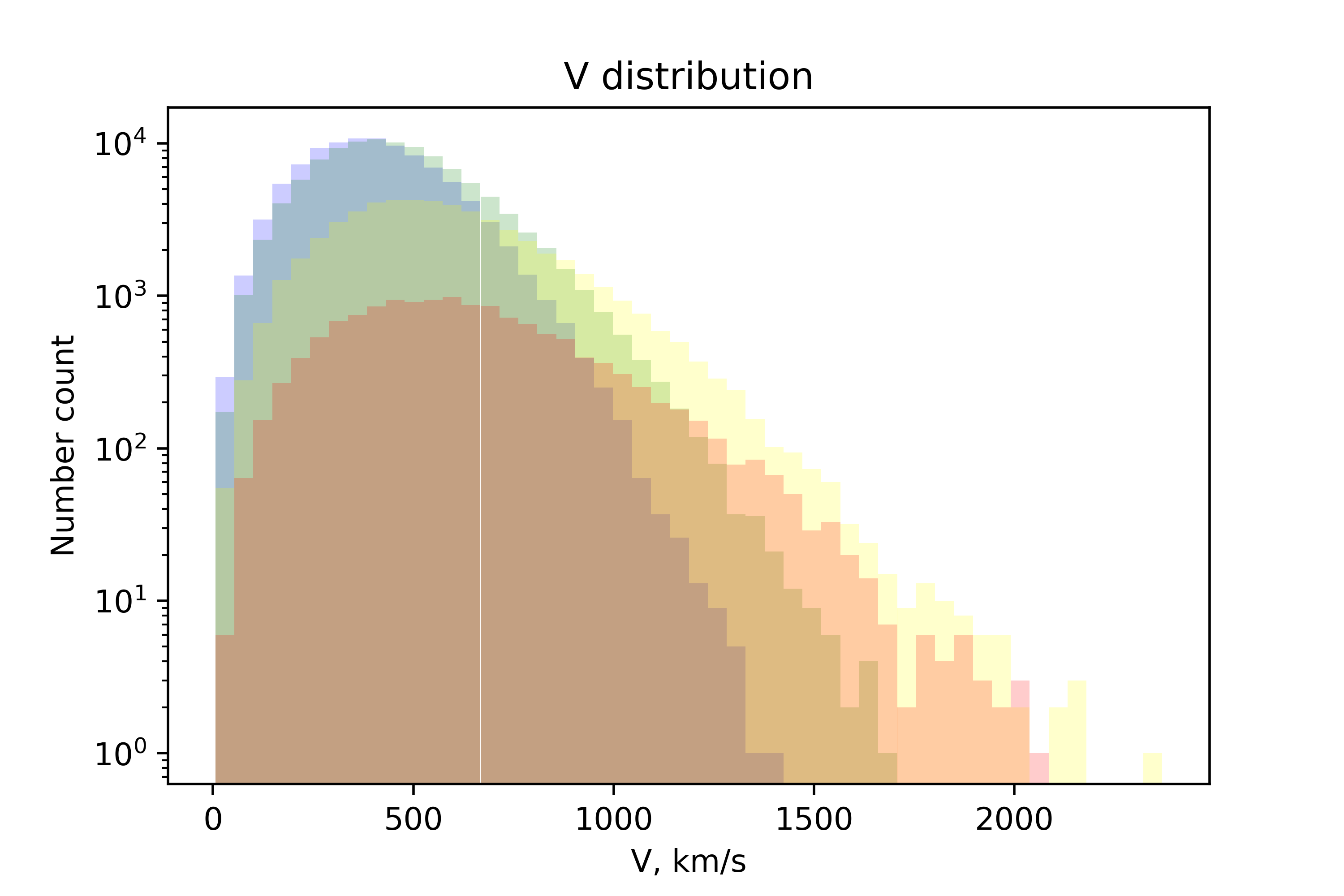}
  \end{center}
  \caption{Distribution of masses (left panel) and velocities (right panel) in different topology classes subpopulations: voids (blue), filaments (green), sheets (yellow) and knots (red).}
\label{figmass}
\end{figure*}

Among 12 topology structure classification schemes compared in \cite{Libeskind17} we are interested in those, which have all 4 classes in their classification of topological structures, there are 6 of them. These schemes are called there T-web \citep{Forero-Romero09}, V-web \citep{Hoffman12}, NEXUS+ \citep{Cautun13}, ORIGAMI \citep{Falck12}, MWSA \citep{Ramachandra2015} and CLASSIC \citep{Hahn07}. For the last of the listed, CLASSIC, we obtained the best score when using ML (see Table 3 for comparison), so we choose it for detailed analysis here. This method is based on linearization of cosmological density field and evaluating the number of eigenvalues of the Hessian of the gravitational potential. Depending on its value a corresponding type of topological structure (void, filament sheet or knot) is assigned to each node. In the catalogues provided in \citep{Libeskind17} for each node its type of structure is coded as "0", "1", "2" and "3" (in web$\_$ID column of the data file), corresponding to voids, filaments, sheets and knots (superclusters) respectively. Beside this number coding we will also use color coding on our graphs, depicting number counts of void population with blue, filament with green, sheets with yellow and knots with red colors. The number count of each population are presented in Figure \ref{fig1b}. The filament is the most populated one. Distribution of mass and velocities in different topology classes subpopulations are given in \ref{figmass}.

\section{Network metrics}
Currently there are definitions for more than hundred different local network metrics of a  node, lot of them are similar and others are less relevant for the Cosmic Web, taking into account the nature of bounds between nodes (halos). For our analysis we have chosen 10 network characteristics evaluated for each node: degree, average neighbour degree (a.n. degree), betweenness, closeness, harmonic, eigenvector (or just eigen) and Katz centralities, clustering coefficient, triangles and squares. All of them were computed with the help of \texttt{NetworkX} package \cite{networkx} for \texttt{Python}. These metrics are widely used when studying social networks\citep{Brandes01}, transport \citep{Holovatch12}, communications and other kind of  networks\citep{Albert01}, including Cosmic Web \citep{Apunevych17}. On the other hand, it occurs that adding more metrics of the node doesn't help in ML problem of predicting the topology. Let us go through definition of each of the characteristics.
\subsection{Definitions}
First let us define the metrics based on the number of neighbours of the node. In this section the term ''distance`` between two nodes will refer to the number of edges separating them. Namely, that the distance between nodes is measured in terms of the number of edges in the shortest path between them. 
\begin{itemize}

\item{\textit{Degree}}. The degree $k_j$ of the node $j$ is one of the basic network metrics and is defined simply as the number of its neighbours, that is, nodes that share a common edge with it. In terms of the adjacency matrix:
$$k_j = \sum_{i=1}^N a_{ij}.$$
Here and below the summation is carried out over all $N$ nodes of the network, if not explained explicitly. 
\item{\textit{Average neighbour degree}} (a.n. degree) $A_n$ is literally, average degree of the neighbours of node $j$, normalized by number of neighbours:
$$A_n(j) = \frac{1}{k_j}\sum_{t \in n(j)} k_t. $$ 
$n(j)$ denotes set of neighbours of the node $j$.
\item{{\it Katz centrality}} \citep{Katz53} is a further, higher order generalization of the node degree: it takes into account not only the number neighbours, but also neighbours of neighbours etc. The number of more distant neighbours is weighted inversely with distance. In terms of the elements of the adjacency matrix $a_{ji}$ it is defined as follows:  
$$C_{\mathrm{K}}(j) = \sum_{l=1}^\infty \sum_{i=1}^N \alpha^l (a^l)_{ji} .$$
Here $(a^l)_{ji}$ denotes element of the adjacency matrix of nodes at distance $l$, which means that its element is 1 if $i$ and $j$ nodes have $l$ elements between them and 0 otherwise. Coefficient $\alpha$ is picked up to keep the sum convergent, but at the same time to take into account as distant neighbours as possible. In this work $\alpha$ is taken to be 0.02, which is a compromise between computational difficulty and depth of characterising network with this metric.
\end{itemize}
Next, let's consider other centrality metrics.
\begin{itemize}
\item{\textit{Betweenness centrality}}. The (normalized) betweenness centrality of a node $j$ tells us the fraction of shortest paths of all paths that go through the node. It is defined by the expression \citep{Brandes01} :
$$C_b(j)= \frac{2}{(N-1)(N-2)}\sum_{s,t=1}^N\frac{\sigma_{st}(j)}{\sigma_{st}}.$$
Here $\sigma_{st}$ is the total number of shortest paths from node $s$ to node $t$ and $\sigma_{st}(j)$ is the number of those paths that pass through $j$. If $s=t$ then $\sigma_{st} = 1$ and if $j=s$ or $j=t$ then $\sigma_{st}(j) = 0$. The metric is 0 if $s$ and $t$ are not connected.
\item{\textit{Closeness and harmonic centralities.}} The (normalized) closeness centrality of a node $j$ is \citep{Brandes01} reciprocal to the sum of distances $d(y,j)$ to all other nodes, to which it is connected ($V(j)$). It is normalized by the number of nodes:
$$C_c(j)= \frac{N-1}{\sum_{y \in V(j)} d(y,j)}, $$
A very similar definition has the harmonic centrality of the node \citep{Vigna14}. It is the sum of the reciprocal of the shortest path distances from all other nodes to $j$:
$$C_h(j)= \sum_{y \in V(j)} \frac{1}{d(y,j)},$$
where $1/d(y,j) = 0$ if there is no path from $y$ to $j$. Unlike closeness, harmonic centrality was computed without normalization.
\end{itemize}
Now, let us introduce network metrics related to clustering.
\begin{itemize}
\item{\textit{Clustering coefficient and triangles}}. Triangles of the node $T(j)$ have very simple definition: it's a number of triangles, formed by edges, that include $j$. Clustering coefficient \citep{Barthelemy10} is, in addition, normalized by the number of possible triangles, if all neighbours are connected. 
$$Cl(j) = \frac{2T(j)}{k_j(k_j-1)}.$$
On the other hand, clustering coefficient is equal to fraction of connections between neighbours to all possible connection between them. Also triangles are natural mix of the clustering coefficient and the degree of a node. 
\item{\textit{Square clustering}} of the node $j$, $Sq(j)$, is computed in a similar way as the clustering coefficient, it is a fraction of squares formed by edges of all possible squares if all neighbours are connected.
\end{itemize}

Finally, \textit{eigenvector centrality} or \textit{eigencentrality} has to be introduced. The eigencentrality is a measure of the influence of a node in a network. Scores are assigned to nodes based on the concept that connections to high-scoring nodes contribute more to the score of the node:
$$C_x(j) = \frac 1 \lambda \sum_{t \in n(j)} C_x(t),$$
where $\lambda$ is some constant.
Or, in terms the adjacency matrix $\mathbf{A}$, the $j$-th component of eigenvector $\mathbf{x}$ of matrix is the eigenvector centrality of the $j$-th node, $\mathbf{Ax} = {\lambda}\mathbf{x}$.
So the node is important if it is linked to other important nodes, and eigencentrality quantifies the measure of importance. It is computed iteratively, and is only defined up to a common factor, so only the ratios of the centralities of the vertices are well defined.

Next, let us compute the correlation coefficients with type (rank) of topology structures and between network metrics themselves.

\subsection{Correlation between network metrics and topology class}
Our aim is to find out how useful the network information itemized in the previous section is, for the study of the topology of the Cosmic web.. Therefore, we are interested in local (individual) values of network metrics for each node. Nevertheless, for interested reader we provide averages of network metrics in Table 1 for networks with different linking length. Recall, that the results, presented below, are obtained for network with linking length $l=2$ $h^{-1}$ Mpc (the third column of the Table 1).

Our first step is to find the Spearman rank correlation coefficient for all the network metrics of the node with the type of topology structure to which the node belongs. Spearman's correlation measures the strength and direction of association between two variables, i.e. whether the dependent one grows if the independent does. In our case, the independent variable $X$ is the code of the topology type structure with possible values "0", "1", "2" or "3", as was introduced in \citep{Libeskind17}, and variable $Y$ denotes one of the 10 metrics introduced above:
$$r_s  = \frac {\operatorname{cov}(X,Y)} { \sigma_{X} \sigma_{Y} }$$
So, we have covariance of these variables in the numerator and product of standard deviations of variables in the denominator.

\begin{figure*}
  \begin{center}
  \includegraphics[width=0.48\textwidth]{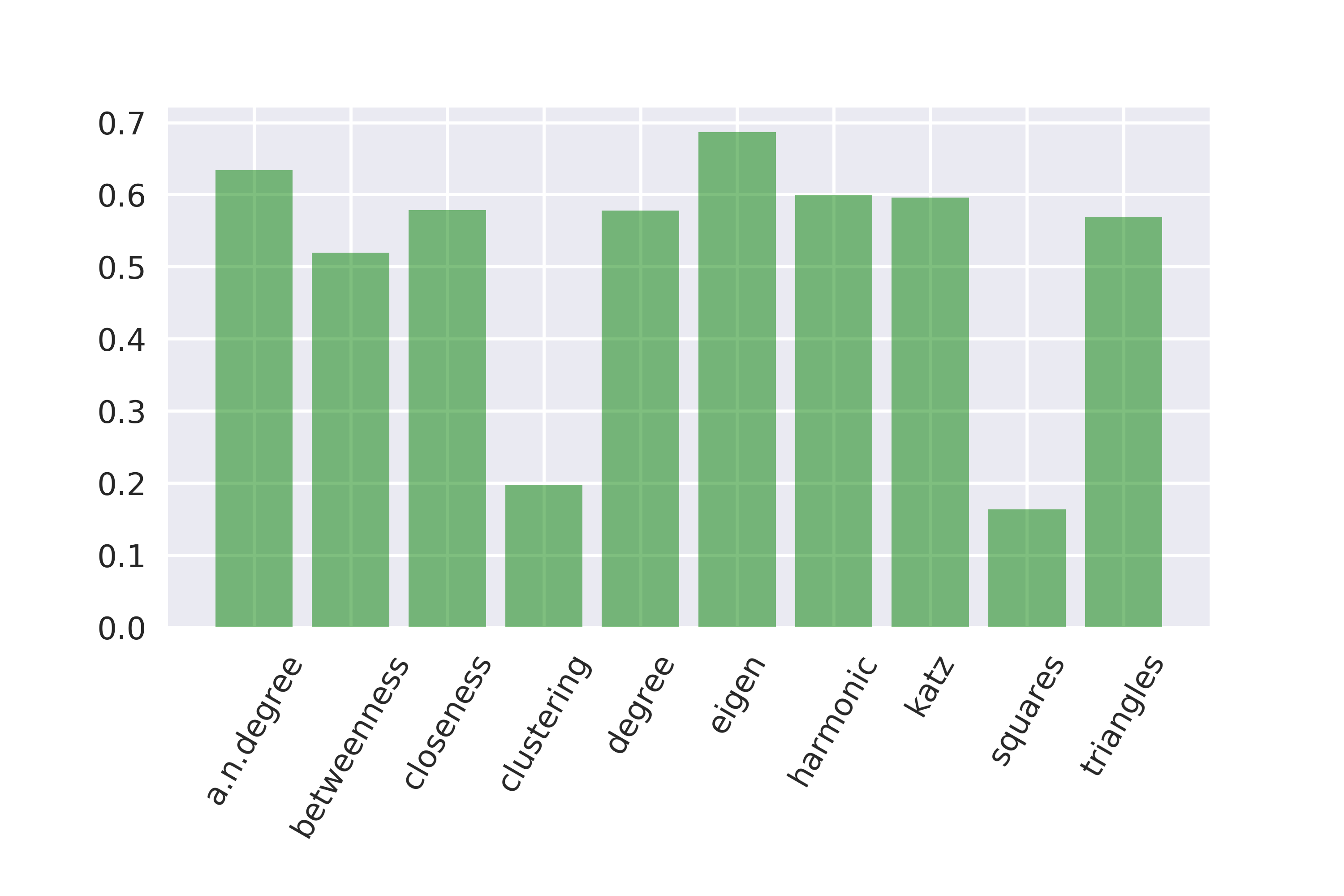}
  \includegraphics[width=0.48\textwidth]{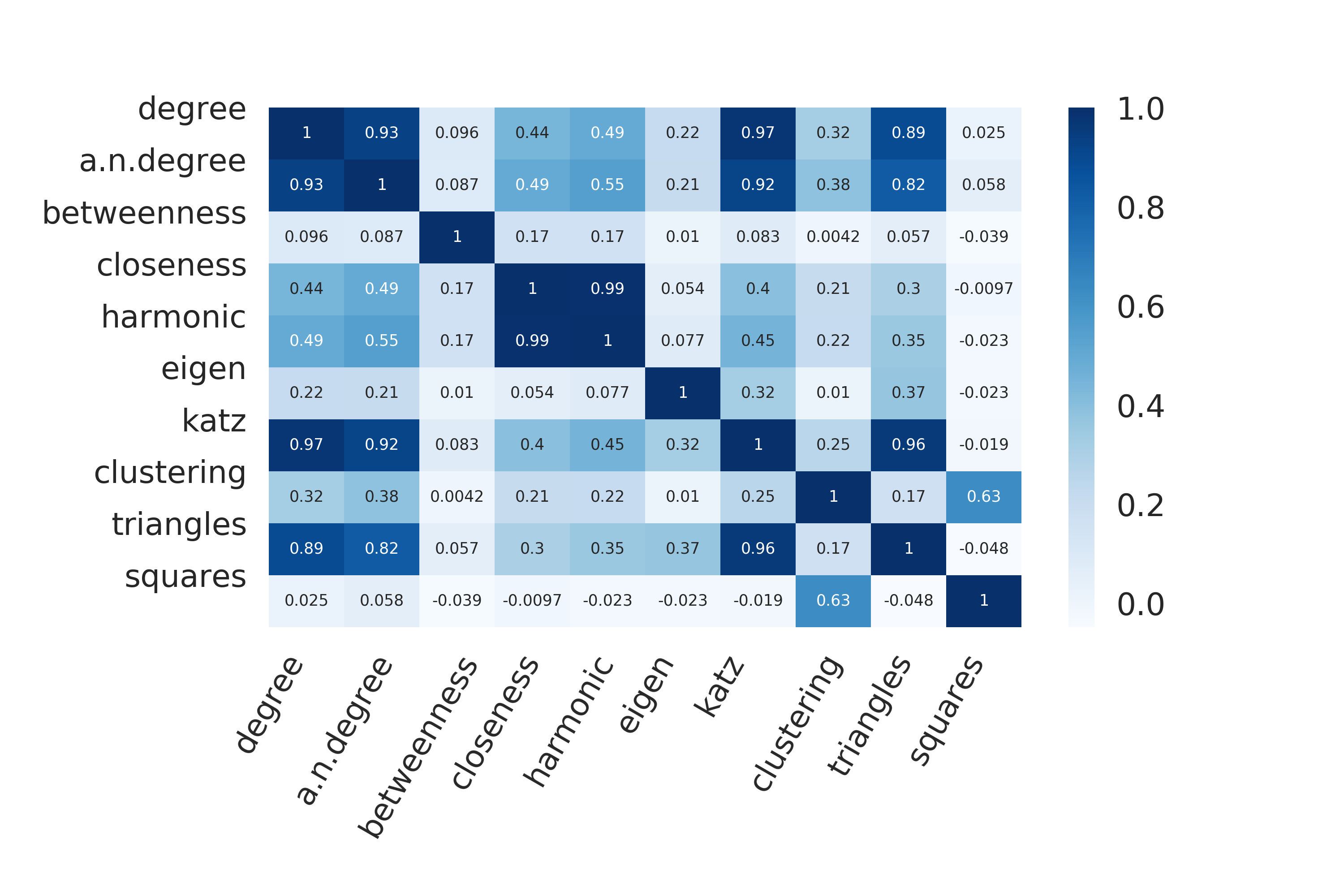}
  \end{center}
  \caption{On the left: Spearman rank correlation between type of topology structure and the network characteristics. On the right: cross-correlations (Pearson coefficient) between different network characteristics}
\label{fig2}
\end{figure*}

Although this is a discrete variable and reflects categorical data, the increment of $r_s$ corresponds to increment of dimensionality of the structure in some sense. Intuitively it is clear that change in $r_s$ corresponds to a change in topological structure, from less dense regions (voids) to more dense regions (knots). The results are given in Figure \ref{fig2}. Note, that the highest rank correlation is with eigencentrality, which means it is an important characteristic for topology of the network, while the lowest is with clustering and square clustering coefficients, indicating that they don't differ too much for populations of different topology structure. 
Differences in Spearman rank correlation coefficients increase with growth of the linking length.

Next we compute the correlations between network metrics themselves. To quantify correlation between continuous values (network metrics) the Pearson
coefficient is used, while the Spearman coefficient serves to find correlation between continuous and discrete (categorical) values as the topology index. Pearson coefficient is computed for all 45 different pairs of metrics and present in a form of the heatmap in Figure \ref{fig2}. This plot shows how correlated different metrics in our network are. One can note, that degree, average neighbour degree and Katz centrality correlate between themselves strongly, which makes sense, as latter two are generalization of the first one. Closeness and harmonic centrality correlate as they have similar definition. The most independent are betweenness centrality and squares. Cross-correlations between different network characteristics shown in the right panel of Figure \ref{fig2} reflect unique features of the network structure. Another example is given by observing that the clustering coefficient has a high correlation with square clustering, while it has a low correlation with triangles and degrees. Such behaviour my be explained by noticing that the clustering coefficient is linear function of triangles, but inverse-quadratic function of degree of a node. A node can have high degree and moderate number of triangles and
hence low clustering. On the other hand, it can have low degree, low number of triangles and low clustering. Also, the squares have low correlation with degree, while triangles have a high one. This may explain the fact that clustering coefficient correlates differently with triangle
and squares.

\subsection{Distributions for different topology class sub-populations}
\begin{figure*}
  \begin{center}
  \includegraphics[width=0.48\textwidth]{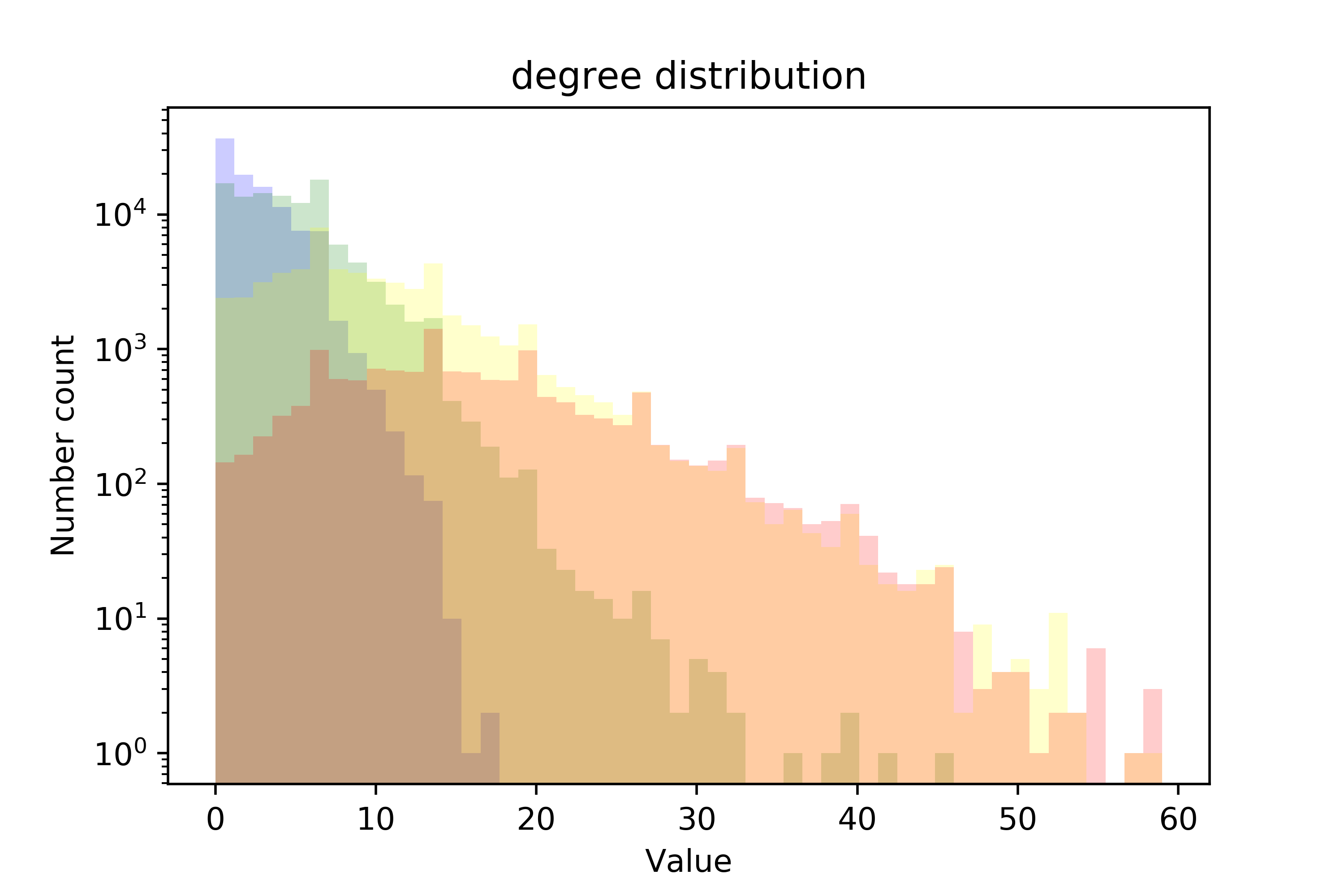}
  \includegraphics[width=0.48\textwidth]{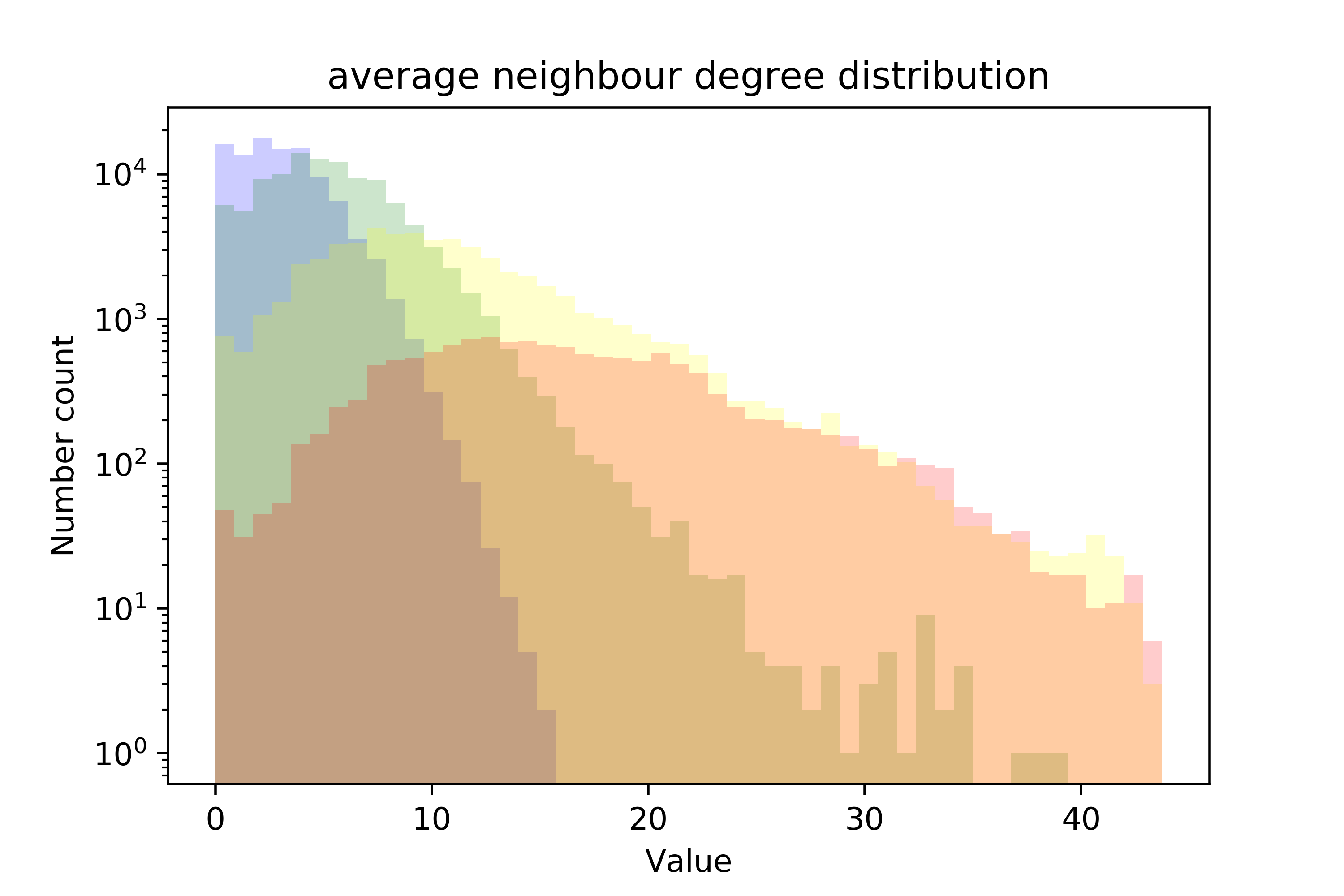}
  \end{center}
  \caption{Number count histogram for degree (left) and  average neighbour degree (right) in voids (blue), filaments (green), sheets (yellow) and knots (red).}
  \label{fig3}
\end{figure*}

\begin{figure*}
  \begin{center}
  \includegraphics[width=0.48\textwidth]{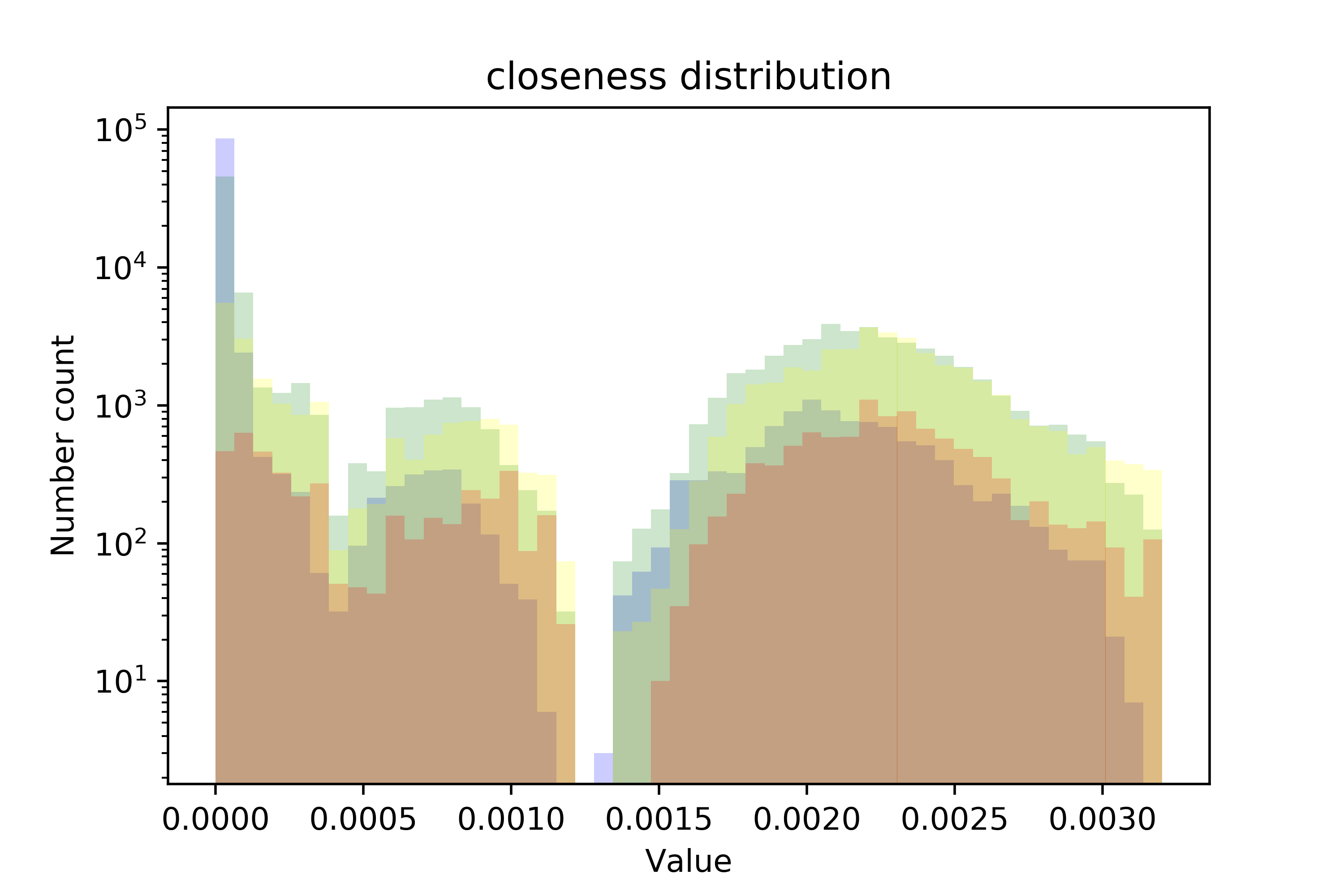}
  \includegraphics[width=0.48\textwidth]{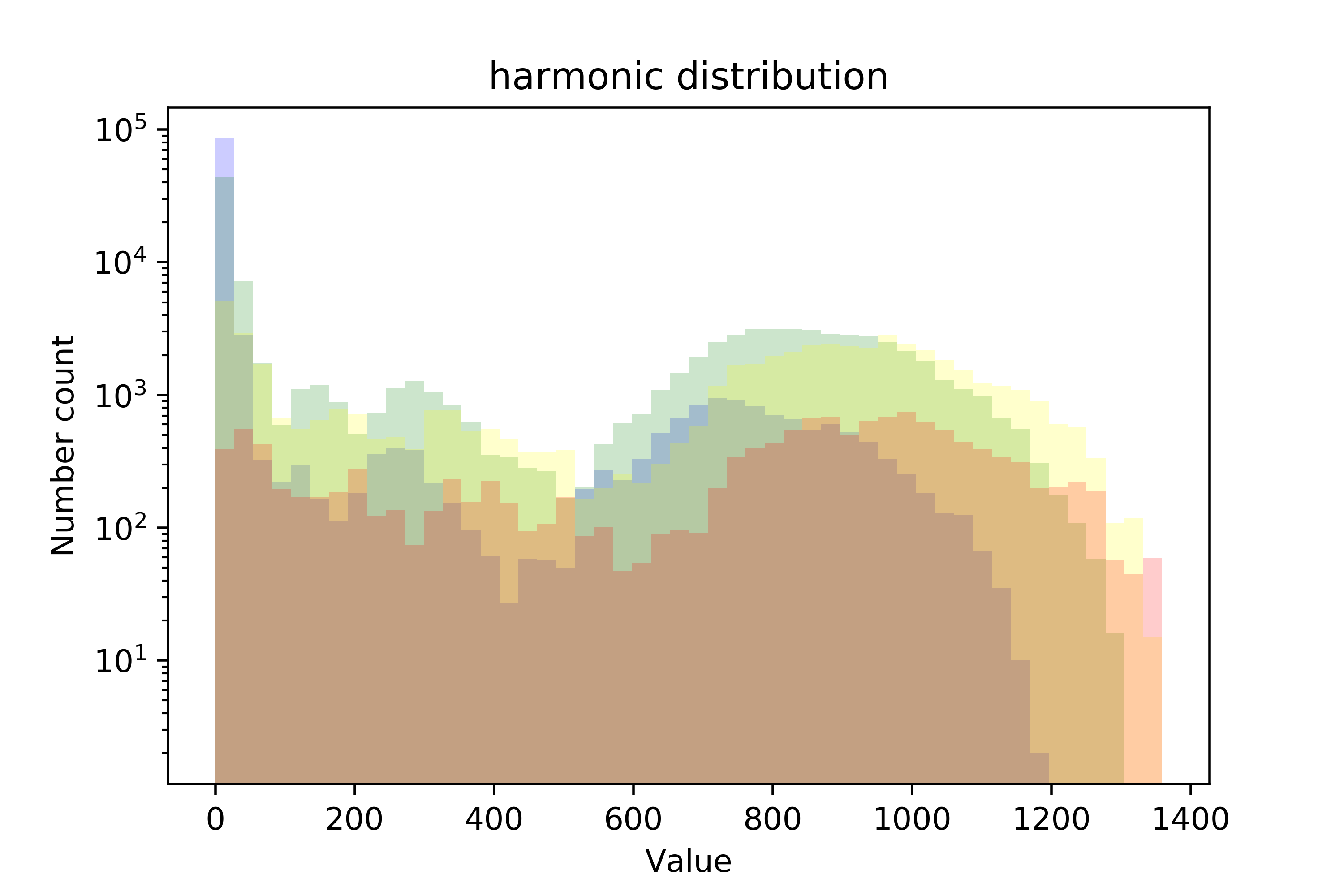}
  \end{center}
  \caption{Number count histogram for closeness(left) and harmonic (right) centralities in voids (blue), filaments (green), sheets (yellow) and knots (red).}
  \label{fig4}
\end{figure*}

\begin{figure*}
  \begin{center}
  \includegraphics[width=0.48\textwidth]{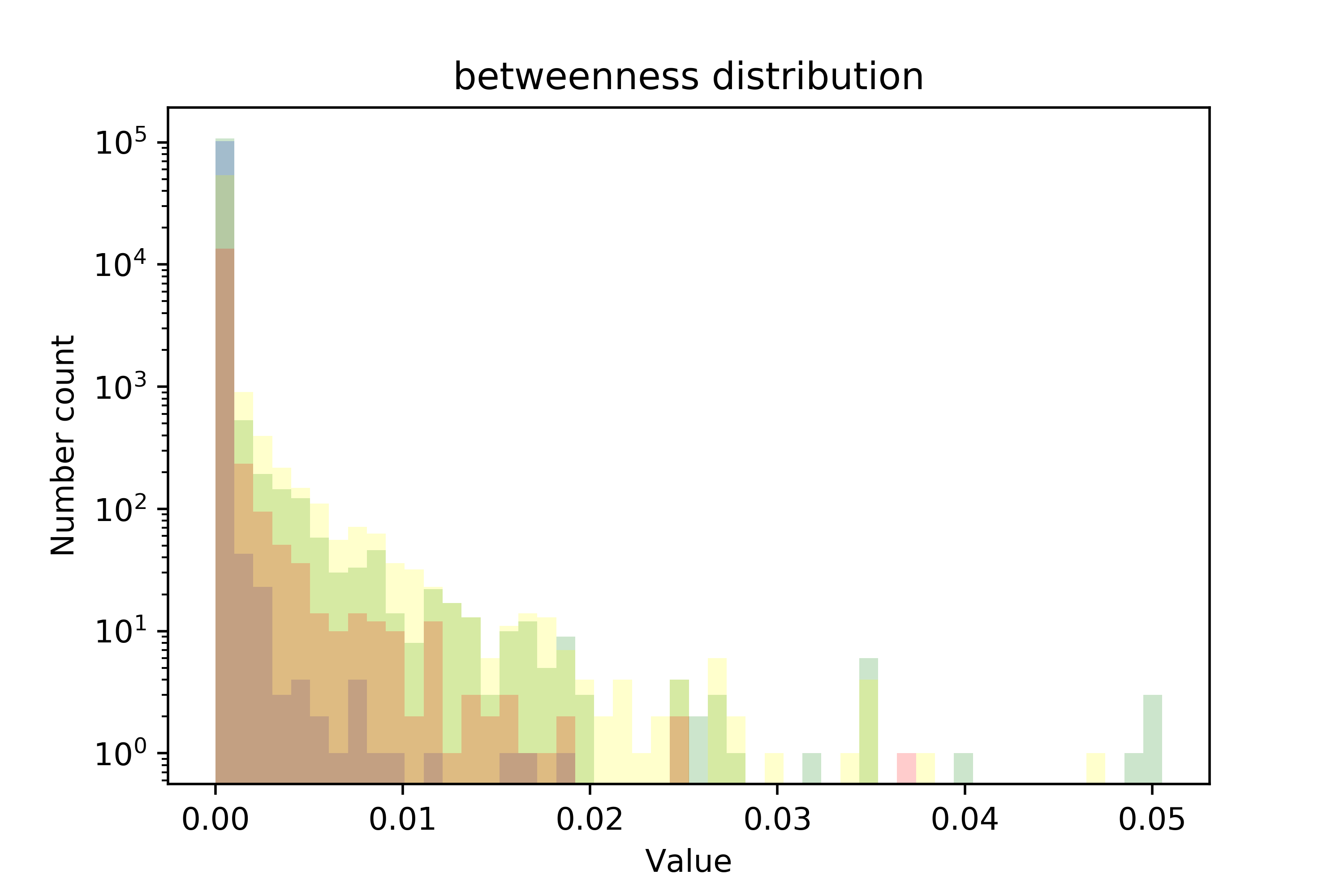}
  \includegraphics[width=0.48\textwidth]{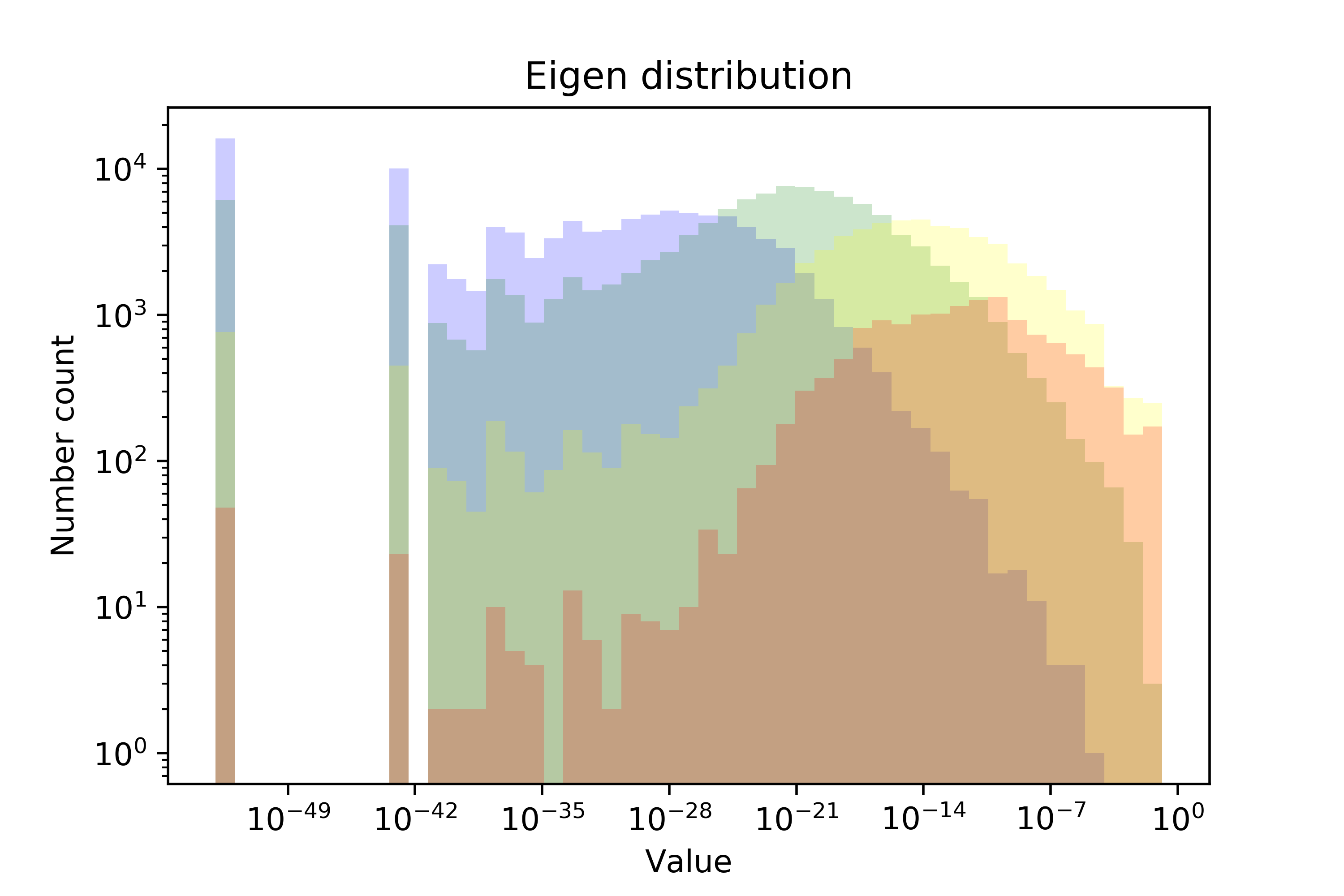}
  \end{center}
  \caption{Number count histogram for betweeness(left) and eigencentrality(right) in voids (blue), filaments (green), sheets (yellow) and knots (red).}
  \label{fig5}
\end{figure*}

\begin{figure*}
  \begin{center}
  \includegraphics[width=0.48\textwidth]{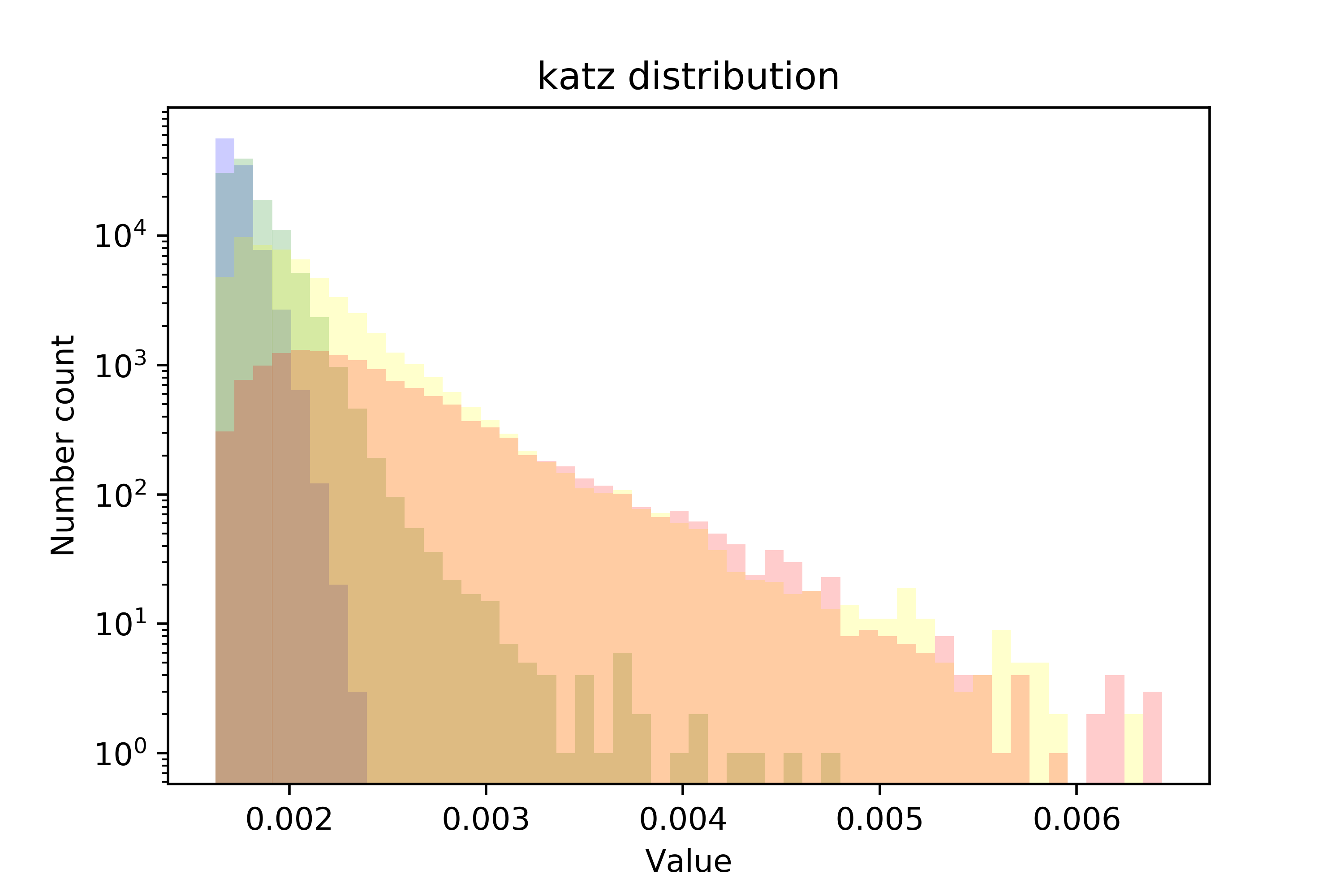}
  \includegraphics[width=0.48\textwidth]{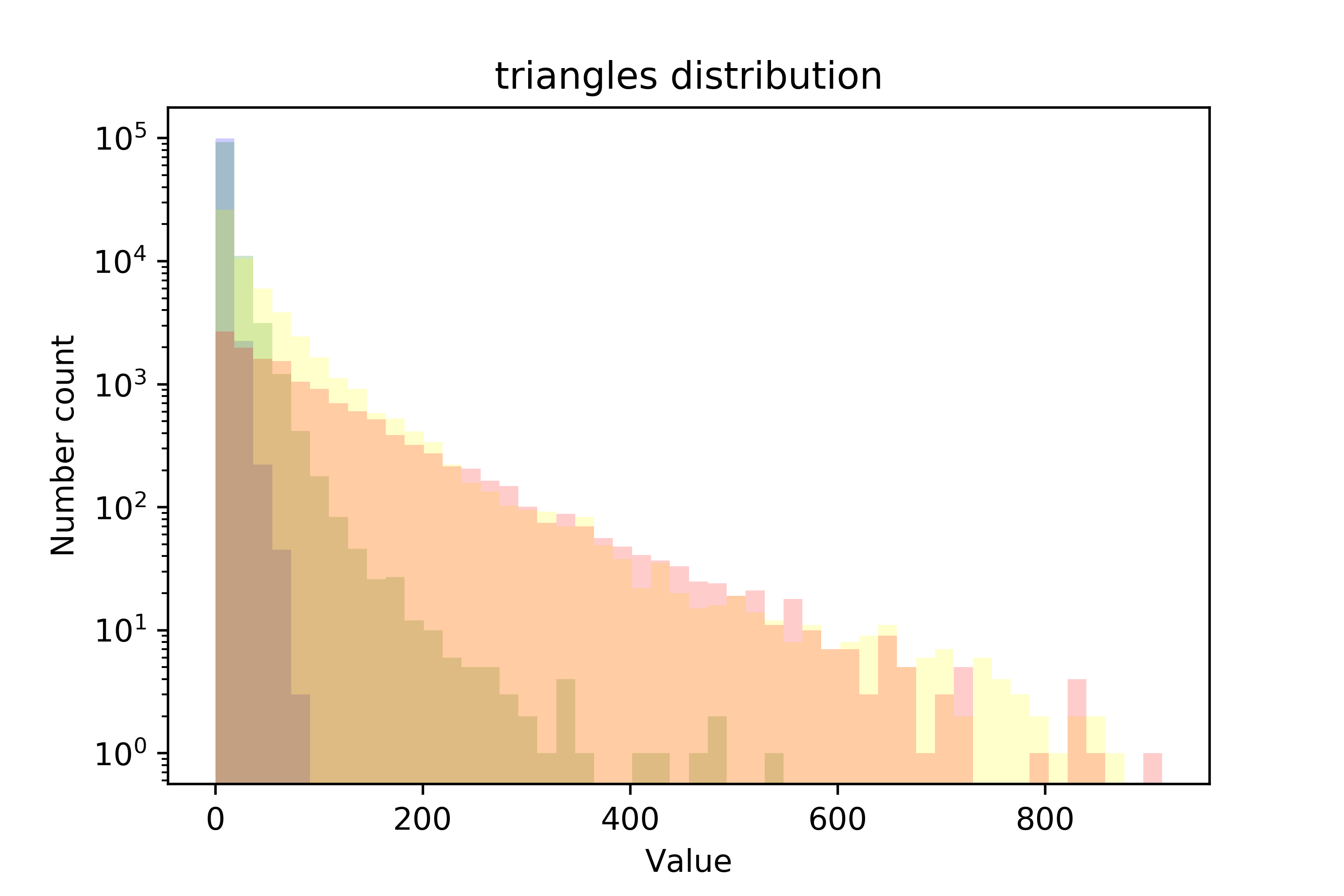}
  \end{center}
  \caption{Number count histogram for Katz centrality (left) and triangles (right) in voids (blue), filaments (green), sheets (yellow) and knots (red).}
  \label{fig6}
\end{figure*}

\begin{figure*}
  \begin{center}
  \includegraphics[width=0.48\textwidth]{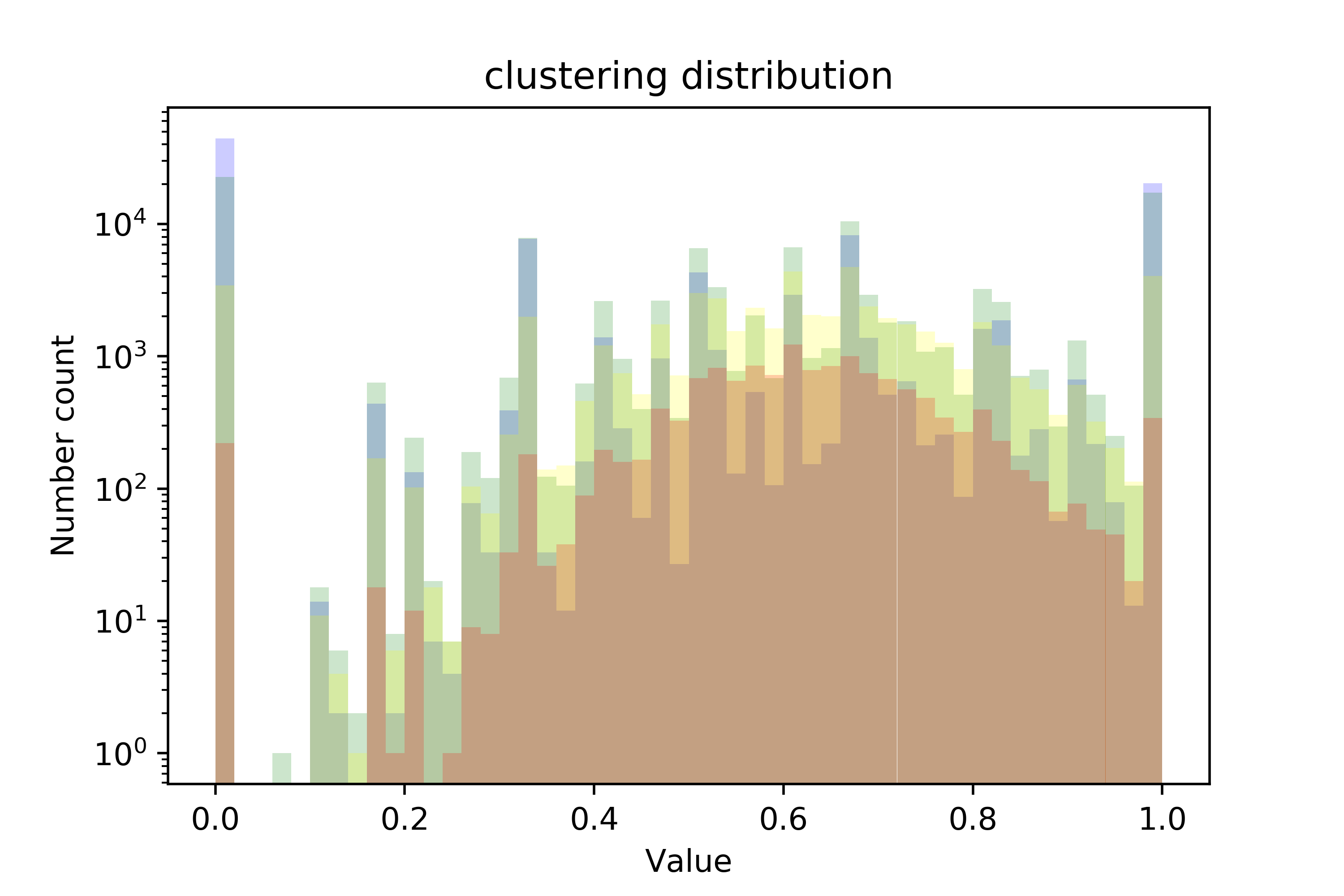}
  \includegraphics[width=0.48\textwidth]{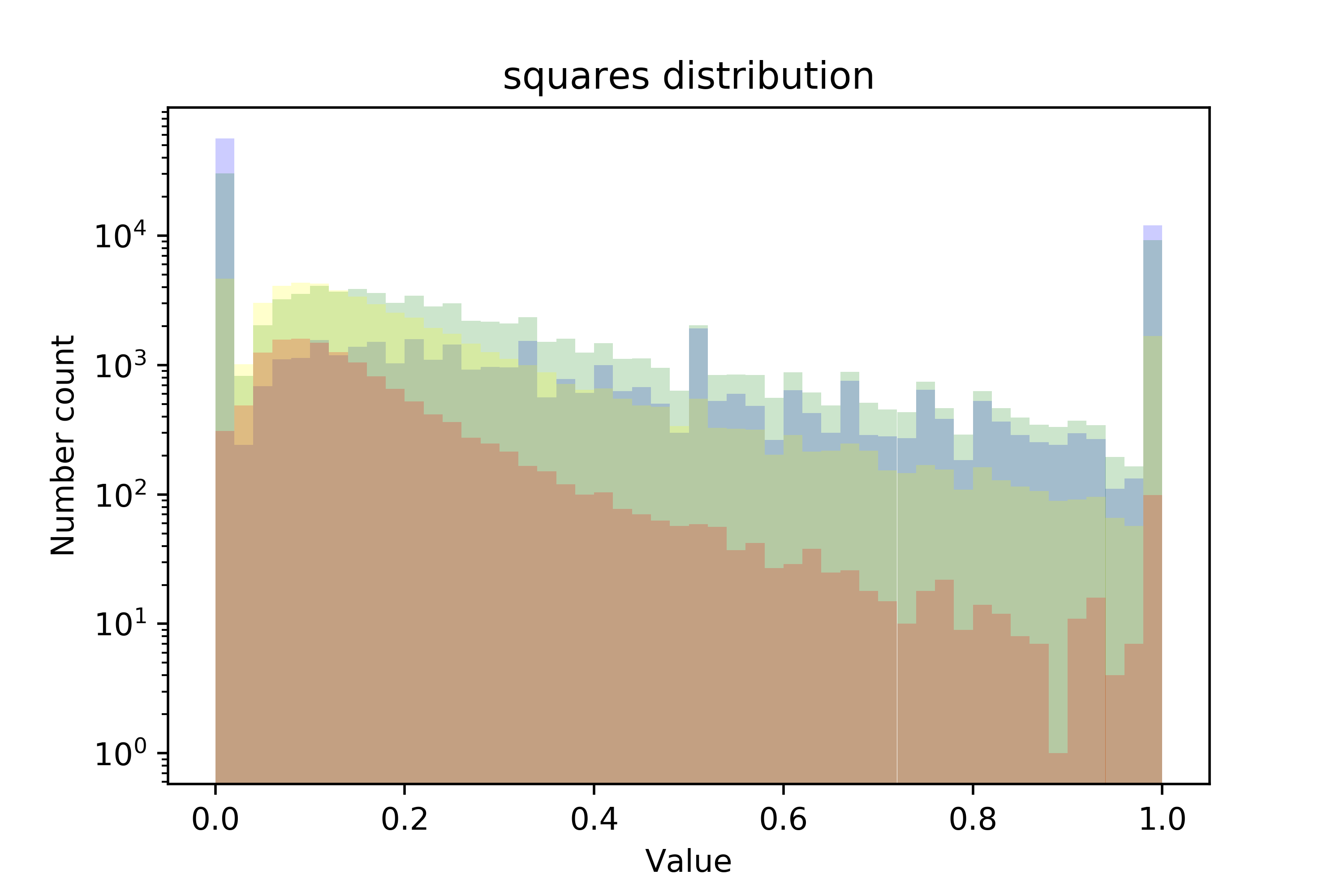}
  \end{center}
  \caption{Number count histogram for clustering coefficient(left) and square clustering (right) in voids (blue), filaments (green), sheets (yellow) and knots (red).}
  \label{fig7}
\end{figure*}

Now lets have a look at the number count distributions of the characteristics for subpopulations of different topology. Results are presented in Figures \ref{fig3}--\ref{fig7}.As expected, the higher the correlation between a given metric and the topological structure, the narrower the distribution and the more well defined the maxima will be. In practice we observe this for degree, average neighbour degree and Katz centrality: the maxima in distributions are well distinguishable, they have different form and skewness in the distributions. However, this rule doesn't hold for the harmonic centrality, which correlates more than degree with the topology index.
Consequently, we distinguish three types of metrics by types of distributions in subpopulations of different topology structures: those that have drastically different distribution for different types of large scale structure (degree, a.n. degree, Katz centrality, triangles), those that have similar distributions in different subpopulations (closeness centrality, clustering coefficient, square clustering coefficient), and those that are somewhere in the middle between previous two groups, having somewhat different distributions (harmonic, eigenvector, betweenness centralities). Distributions of degree and clustering coefficient we obtained have comparable distribution with those in \citep{Apunevych17}. It is also interesting to point out that clustering coefficient seems to have distribution, indifferent to type of topology structure, while it was shown in \citep{Apunevych17} that distributions of the color index and stellar mass of galaxies as nodes are different for populations with different clustering coefficient. 

There are two metrics, that have clearly low correlation with topology index: clustering
coefficient and squares. Both of them have similar distributions in different subpopulations. On the other hand, there are those which have similar distributions and relatively high correlations (closeness and harmonic centralities). We believe that relatively high correlation of those can be explained by high peaks in distributions for void and filament subpopulations for 0 values of centralities. Note, that number count axis is given in logarithmic scale.

\section{Machine learning and predictive power of network characteristics}
The general purpose of using ML in this work is to reproduce topological classification of the Cosmic Web obtained within certain method having just a few characteristics (predictors) for each halo. We have used the following information about halos: spacial coordinates, masses and peculiar velocities, which are fully available with good accuracy in synthetic data. Based on spacial coordinates we found 10 network metrics for each halo, which (together with peculiar velocity and mass) are the predictors in our ML models.

After having tried several ML techniques to compare their predictive power (see Table 2 with comparison of prediction scores), it turns out, that the extreme gradient boosting decision trees method (with realization via \texttt{xgboost} library on \texttt{Python}) of classification is the most efficient to predict the topology structure type of nodes. 
\begin{table}
\begin{tabular}{|c|c| } 
 \hline
  ML technique & Prediction score \\
  \hline 
  Multi layer perceptron & 0.523 \\
  k nearest neighbours & 0.598 \\ 
  RandomForest decision trees & 0.653 \\ 
  xgboost & 0.700 \\ 
 \hline
\label{table3}
\end{tabular}
\caption{Comparison by prediction score different ML techniques}
\end{table}
One can find detailed description of this method in \citep{Chen17}, while here we provide a short sketch.
\begin{figure}
  \begin{center}
  \includegraphics[width=0.48\textwidth]{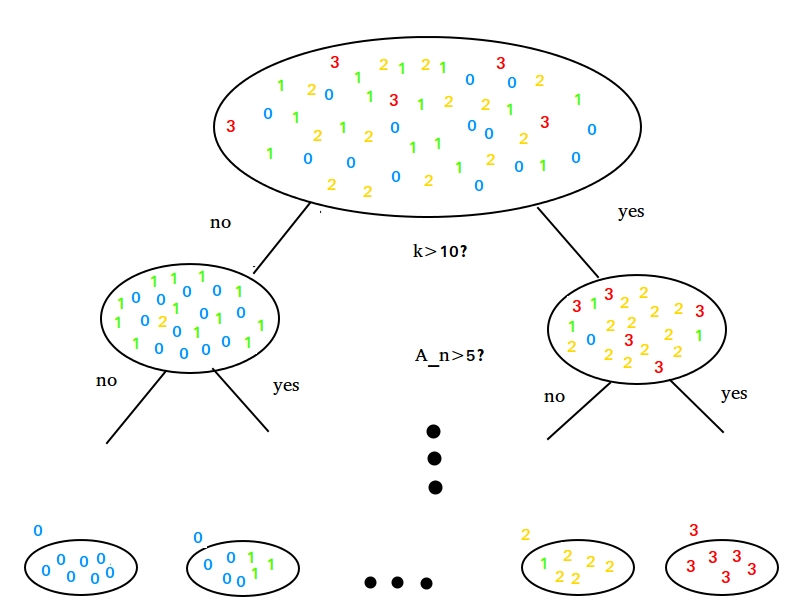}
  \end{center}
  \caption{An example of how a decision tree can be formed in our problem. The ``depth'' (number of splits) of a tree can be controlled, it is a hyper-parameter of the method. Value of the predictor, which splits the sample at each step is chosen to maximize the ``entropy'' after splitting, that is to have maximally different subsets at the next step. Final subsets are pre-scripted to one of the classification categories. The prediction is given based on which bucket the instance belongs to after the same series of questions about its characteristic.}
  \label{tree}
\end{figure}
As any other ML classifier, xgboost uses the vector of predictors (also called ``features'') $X$ (10 network metrics plus mass and velocity of the halo in our case) to produce prediction $\hat{y}$ of values of the target variable $y$ (index of topology structure in our case). Goal of any ML techniques is to "train" a model (function) $F$ to predict values of the form $\hat{y}_i = F(X_i)$ for $i$-th  instance (halo in our case) by minimizing the loss function $L(\hat{y}, y)$ of the prediction:
$$L(\hat{y}, y) = \sum_{i=1}^M l(y_i, \hat{y}_i)$$
where summation is over training set of size $M$ (in our case $M=0.9N$) and $l(y_i, \hat{y}_i)$ function that measure the difference between prediction and target variable value for each instance. For regression model it can be mean squared error, 
$$L(\hat{y}, y) = \tfrac{1}{M}\sum_i^M (\hat{y}_i - y_i)^2.$$ In (our) case of classification model, it is usually taken as a cross-entropy loss function, 
$$L(\hat{y}, y) = -\tfrac{1}{M}\sum_i^M \left(y_i \log p_i + (1-y)\log(1-p_i) \right),$$ where $p_i=p_i(\hat{y}_i)$ is the probability given by model to predict the correct category $y_i$ of instance $i$ based on its features $X_i$.

The family of decision tree techniques uses an idea of building the classification tree, in which ``branching'' occurs when splitting the set of instances (halos) by some specific predictor values at each step. After a number of such steps one obtains a subsets, instances in which belong to (almost) the same classes. The ensemble of such trees (with prediction function $f_k(X_i)$ for $k$-th tree), which are weak classifiers, is used to obtain the final prediction. The final prediction is the result of ``voting'' of predictions from the each tree. The objective function (which is to be optimized in the model) can be written as:
$$\mathcal{L} = \sum_i^M l(y_i, \hat{y}_i) + \sum_k \Omega (f_k), \quad $$ 
where $\hat{y}_i = \sum_k f_k(X_i)$ and summation $\sum_k$ is over ensemble of decision trees. Function $l$ has to be differentiable. Last term, $\sum_k \Omega (f_k)$ is so-called regularization term. It penalizes the complexity of the model and increases with sum of squares of predictors weights in functions $f_k$. We show a sketch of a possible decision tree for our problem in Figure \ref{tree}.

Gradient boosting of decision trees is a way to enhance the prediction of the tree ensemble with the method of gradient descent. The object function is built iteratively, at each step adding function $f_t$ that improves our model the most. On the $t$-th step:
$$\mathcal{L}^{(t)} = \sum_i^N [l(y_i, \hat{y}_i) + g_i f_t(X_i) + \frac{h_i}{2}f_t^2(X_i)] + \sum_k\Omega (f_k), $$ 
where gradient statistics on the loss function $g_i$ and $h_i$ are improved at each iteration step. Usually, a few hundreds of iterations is made when training such model.

\begin{figure*}
  \begin{center}
  \includegraphics[width=0.48\textwidth]{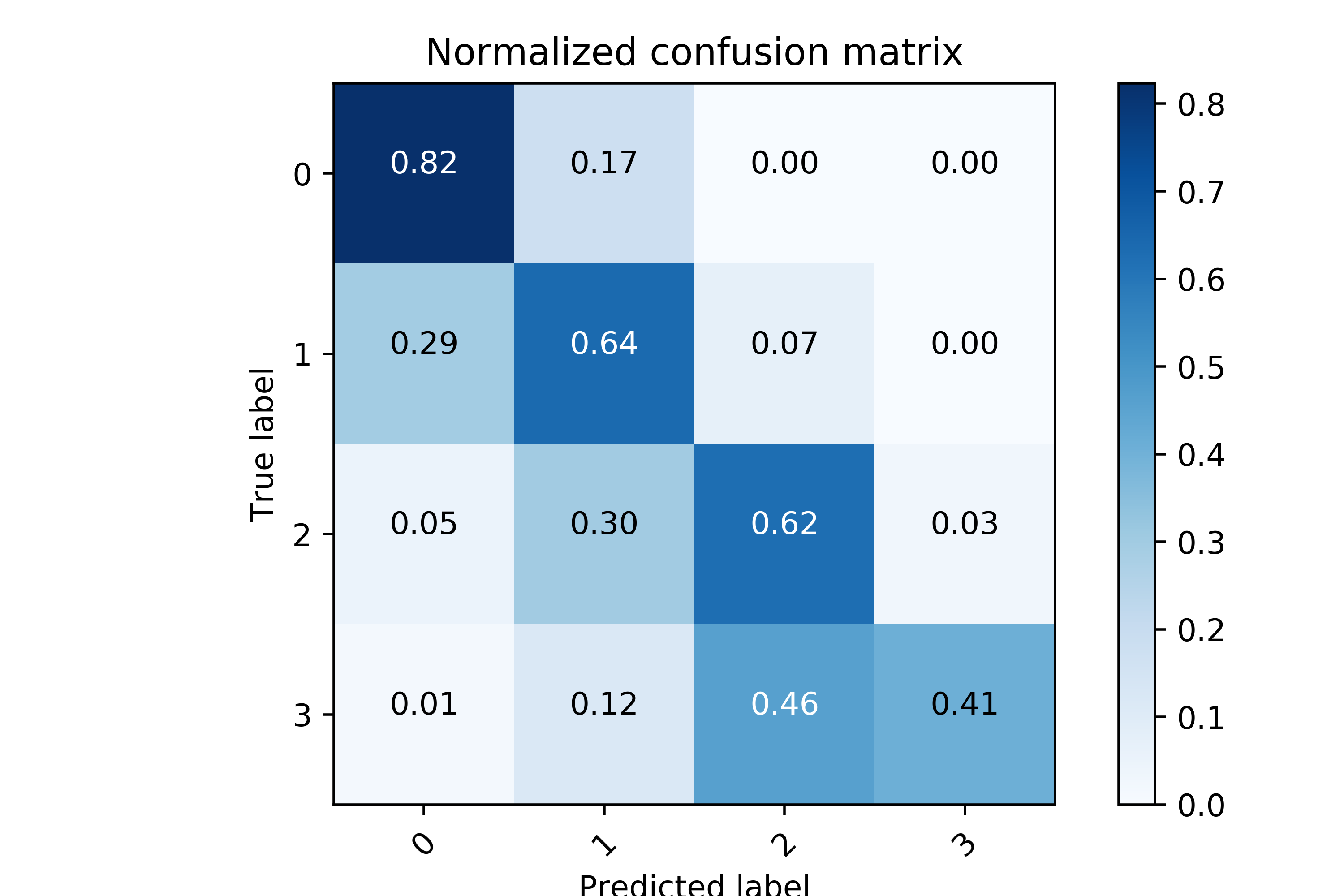}
  \includegraphics[width=0.48\textwidth]{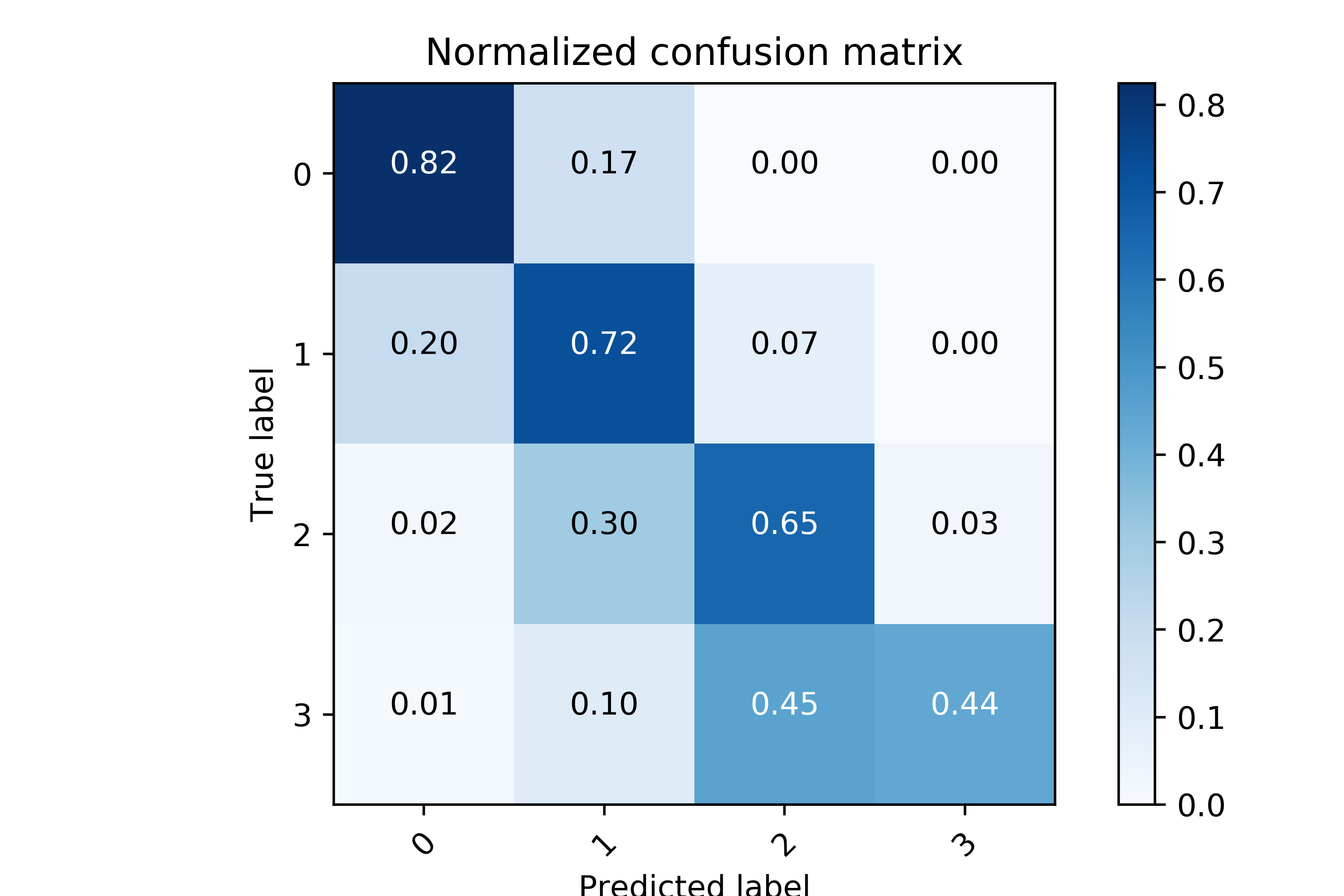}  
  \end{center}
  \caption{Confusion matrix of prediction with xgboost method, on the left: for 2.0 $h^{-1}$ Mpc, on the right: for 2.4 $h^{-1}$ Mpc}
  \label{fig8}
\end{figure*}

Applying described above ML method gave us following results. Even after running through a set of hyper-parameters of the method (number of trees and depth of each), \texttt{xgboost} yields best prediction score only $~70\%$. See Figure \ref{fig8} (left panel) for the normalized confusion matrix of the predictions that was obtained. The score for void population is the best ($82\%$), meaning that this method fits the best for distinguishing void population from others in galaxy distribution, while distinguishing between superclusters and other types of structures with such a method is really unsatisfying: it will predict correctly in less than half of cases. Prediction rates for other classification algorithm are given in Table 2.

We also ran the same method for a network built with the linking length $l = 2.4$ $h^{-1}$ Mpc. The results are given in Figure \ref{fig8} on the right panel. They are slightly better: prediction score is $~72\%$, some entries of confusion matrix have improved. This may be a result of the fact, that the larger linking length yields the larger range of metrics values, which in its turn, allows the ML model to give more accurate prediction. But one should remember, that larger linking length requires more computational resources for evaluation all the metrics.

\begin{table}
\begin{tabular}{|c|c| } 
 \hline
 Classification algorithm & Prediction score \\
  \hline 
  T-web & 0.510 \\
  V-web & 0.551 \\ 
  NEXUS+ & 0.617 \\ 
  ORIGAMI & 0.509 \\ 
  MWSA & 0.624 \\ 
  CLASSIC & 0.700 \\ 
 \hline
\label{table2}
\end{tabular}
\caption{Prediction rate for different topology classification algorithm}
\end{table}
\section{Discussion}
In this work we have applied network analysis to a $\Lambda$CDM cosmological simulation. We computed 10 network metrics on the halo distribution of a publicly available simulation  \citep{Libeskind17}. This simulation is useful because it has been used by multiple groups as benchmark for large scale structure quantification. For each metric the correlation with each node's topology class was computed as well as the cross-correlation between themselves. 
As a result, it is possible to identify which metrics are more important for topological classification. These are degree, average neighbour degree, eigencentrality and harmonic centrality. We also examined distributions of values of those metrics for subpopulations of each type of structure. For some the difference is visible to the naked eye and this gives us a hint that these metrics may be applied to the study of  the large-scale distribtuion of matter.

We studied networks built with different linking length in the range 1.6 -- 2.4 $h^{-1}$ Mpc. The results described here remain mainly the same within the range, showing that there is no any special scale in this range in the Cosmic Web.

We should mention again, that the most interesting would be comparison our results with
those obtained in \cite{HongDei15} and \cite{Apunevych17}. In \cite{HongDei15} authors also linked the topology formed by the dark matter halos of the cosmological simulation (Millennium) with its network metrics: degree centrality, betweenness centrality and clustering. They also studied the distributions of these metrics within the same topology subpopulations. The differences in our results are natural, as we analysed different simulations and the topology structures where defined in another way. In \cite{Apunevych17} authors studied the correlations between network metrics and properties of the node, real galaxies from the COSMOS catalogue in their case. There are distribution histograms there too, for populations of the different redshifts $z$. One can see resemblance of those with our histograms for degree, closeness centrality and clustering. This means that network built on halos of dark matter and galaxies of baryon matter have similar properties.

Another part of the work was applying xgboost ML technique to predict the topology structure initially assigned with the CLASSIC method (in \citep{Libeskind17}). The results shows that, unfortunately, combination of the network analysis and machine learning can not be reliable tool for defining topology structures (voids, filaments, sheets and superclusters) having in background only coordinates, masses and velocities of nodes (halos). The average prediction rate is only $~70\%$ for linking length $l=2$ $h^{-1}$ Mpc and $~72\%$ for $l=2.4$ $h^{-1}$ Mpc. Knots and sheets are most confused with our set of feature variables and the ML techniques applied. Having less of them in the initial classification will provide a better prediction rate. However, we can also see, that some structures are more distinguishable than others, for example, predicting voids has rate of $~82\%$. 

All this brings us to the next steps of network analysis of the Cosmic Web, which can be analysing the dynamics of network, (through dynamics of network metrics, for example) built on the Cosmic Web, with the cosmological evolution of the large scale structure of the Universe.

\section*{Acknowledgments}
This work was supported by the State Found for Fundamental Research of Ukraine under the project F76 and the project of Ministry of Education and Science of Ukraine ``Formation and characteristics of elements of the structure of the multicomponent Universe, gamma radiation of supernova remnants and observations of variable stars'' (state registration number 0119U001544). NIL acknowledges financial support from the Project IDEXLYON at the University of Lyon under the Investments for the Future Program (ANR-16-IDEX-0005). NIL also acknowledge support from the joint Sino-German DFG research Project ``The Cosmic Web and its impact on galaxy formation and alignment'' (DFG-LI 2015/5-1). Authors are thankful to Dr. R. de Regt and Dr. S. Apunevych for inspiring discussions. We also thank the anonymous referee for the useful comments to the first version of the manuscript.


\end{document}